\documentstyle[preprint,aps]{revtex}
\tightenlines
\title{Temperature Dependence of Magneto Current in Spin Valve Transistor: A phenomenological Study}
\author{Jisang Hong and P. S. Anil Kumar}
\address{Max-Planck-Institut f{\"u}r Mikrostruckphysik, Weinberg 2, D-06120 Halle, Germany}
\begin{document}
\maketitle
\begin{abstract}
The temperature dependence of magneto current in the spin spin valve 
transistor system is theoretically explored based on phenomenological 
model.
 We find that the collector current strongly depends on the 
 relative orientation of magnetic moment of ferromagnetic metals due to spin mixing effect. For
 example, the collector current is decreasing in the parallel case 
 with increasing temperature, and it is increasing in anti-parallel configuration. We then obtain decreasing magneto current with increasing temperature. The result accords with the experimental 
data in qualitative manner. This phenomenological model calculations 
suggest that spin mixing effect may play an important role in the spin 
valve transistor system at finite temperature.
\end{abstract}
\newpage
\newcommand{\eb}{\begin{eqnarray}}
\newcommand{\ee}{\end{eqnarray}}
\section{Introduction}
After discovery of giant magneto resistance (GMR) \cite{GMR} in magnetic 
multilayer structure, it has been extensively studied in relation to the 
ferromagnetic tunneling junction. In addition to the traditional 
ferromagnetic junction structure, a new type of potential 
magneto electronic device, so called spin valve transistor, is 
suggested \cite{device}. In the conventional ferromagnetic 
tunneling junction structure, spin dependent transport property of 
electron near 
the Fermi level has been explored. 
But, spin valve transistor (SVT) has different structure \cite{structure}
compared to the conventional ferromagnetic tunneling junction system. 
In a SVT 
structure, electrons injected into the metallic 
base from one side of transistor (emitter side)  pass through the spin 
valve 
and  reach opposite side (collector side) of transistor. While these 
injected 
electrons are passing across the metallic base  they are above the Fermi 
level. These become {\it hot} electrons in spin valve transistor.

The transport property of hot electron may be different from
that of Fermi electron. For instance, spin polarization of Fermi electron
mainly depends on the density of states at Fermi level. However, the spin 
polarization of {\it hot} electron is related to the density of unoccupied
 states above the Fermi level. One can possibly interpret the spin polarization
  of hot electron in terms of inelastic mean free path. 
In ferromagnetic materials, clearly mean free path of probe beam electron 
is spin dependent. For example, Pappas {\it et al} \cite{Papps} measured substantial spin 
asymmetry in the electron transmission through ultrathin film of Fe 
deposted on Cu(100). This implies that understanding of spin 
dependence of the inelastic mean free path is essential to the 
interpretation of the information obtained from spin polarized probe. 
 Most of cases, energy of 
probing beam electron, roughly speaking,  ranges from several eV above the Fermi 
level, and experimental data are interpreted in terms of Stoner 
excitations. Interestingly, in relation to the hot electron transport
 property,  substantial scattering contribution 
from spin wave excitations at low energy was reported in ferromagnetic 
Fe \cite{Fe} experimentally, and also theoretical calculations 
\cite{and,also} present the same results. Based on these results, we believe that spin polarization 
of hot electron at low energy is strongly influenced by spin wave 
excitations. Nevertheless, transport property of hot electron is not fully understood at very low energy
regime at finite temperature. So, it is necessary to probe the temperature dependence of the hot electron transport property more 
explicitly at low energy (for example, within 1 eV range from 
the Fermi level) in relation to the spin valve transistor.  

In spin valve transistor structure,  Jansen {\it et al} reported very 
interesting experimental measurement at finite temperature \cite{finite}. 
They measured collector current across the spin valve changing the 
relative orientation of magnetic moment at finite temperature. 
Surprisingly, they obtain that the collector current has very different feature at finite temperature strongly depending on the relative orientation of magnetic moment. 
The collector current in parallel case is increasing up to 200 K and decreasing beyond that temperature regime. On the other hand in anti-parallel case, the current is increasing up to room temperature. We believe that scattering strength increases with temperature T in 
ordinary metal. This implies that any thermally induced scattering process enhances the 
total scattering. One then  expects that measured current will be decreasing
 with increasing temperature T in any configuration. As the authors of Ref.
  8 
commented, the increase of collector current with temperature T may not 
be related to the ordinary scattering events in the metallic base. Two different 
mechanisms are suggested by the authors of Ref. 8. One is the
 spatial distribution of Schottky barrier height. Authors of Ref. 8
 claim that this may explain the increase of collector
 current in both configurations up to 200 K because more electrons can 
 overcome the Schottky barrier height at collector side with increasing 
 temperature T. However, beyond that temperature regime collector current 
 in parallel case is decreasing while collector current in anti-parallel 
 case is still increasing. Furthermore, this mechanism is not related to any spin dependent property, 
 except for the absolute magnitude of collector current. Therefore, authors of Ref. 8 attribute measured temperature dependence of magneto current to the spin-mixing effect.
 Basically spin mixing is spin-flip 
process  by thermal spin wave emission or absorption at finite 
temperature \cite{flip}. For example, minority electron can flip its 
spin by emitting thermal spin wave, and then it goes into 
spin up channel. In this paper we mainly explore the temperature 
dependence of 
magneto current due to thermal spin wave emission and absorption. 

\section{Phenomenological Model}
For the sake of argument, we assume that spin 
valve has $N/F/N/F/N$ structure where $N$ denotes normal metal, and $F$ represents 
 ferromagnetic metal assuming the same material. Suppose that the same number of spin up and spin 
down electrons with the number $N_0$ are prepared, respectively. In SVT 
structure, the energy of injected electron on the top of Schottky barrier 
at emitter side is around 0.9 eV relative to the Fermi level of metallic base. Here, 
We assume that all source electrons have the same energy. When these 
electrons penetrate magnetic layer certain fractions of source electrons 
will be lost by attenuation factors. We introduce phenomenological 
parameter $\gamma_{M(m)}$ to describe that for majority (minority) spin 
electrons. With initial $N_0$ source 
electrons it is assumed that $N_0\gamma_{M(m)}$ electrons pass the 
ferromagnetic layer if they are majority (minority) spin electrons. 
This phenomenological  parameter $\gamma_{M(m)}$ is related to the spin polarization of hot electrons. 
One should note that spin polarization of hot electrons enters in the 
spin valve system, not that of Fermi electrons.  
There is an example for hot electron spin polarization of  Co \cite{Co}
at very low energy (roughly speaking, 1 eV above the Fermi level). However, spin
polarization of hot electrons, especially for low energy regime, is not clearly understood 
either in theoretically or experimentally at finite temperature. The 
central issue of this paper is in understanding of temperature dependence 
of magneto current by thermal spin wave emission or absorption. 
If one is interested in the absolute 
magnitude of the collector current  one obviously needs to take into 
account many spin dependent scattering events as well as spin 
independent scattering processes. In addition, one also has to 
consider angle dependence \cite{angle} even if electrons have enough energy
to overcome the collector barrier. When
we explore temperature dependence of magneto current we do not include 
any spin and temperature independent 
process  even if it has spin dependent property because all these 
factors are not relevant to the temperature dependence of magneto current.
                                       
One should note that the issue here is the temperature dependence of 
magneto current by spin mixing effect at finite temperature. In that spirit
we 
suppose that spin-flip probability, expressed as $P(T)$, by thermal 
spin wave emission or 
absorption  at finite temperature is proportional to $T^{3/2}$ by virtue
of fact that the number of spin waves at finite temperature are proportional 
to $T^{3/2}$. Without spin flip process, assuming parallel 
configuration, 
the current from spin-up electrons (source electrons with number $N_0$) 
is $N_0\gamma_M^2$, and the current from spin-down electrons is 
$N_0\gamma_m^2$. In the anti-parallel case, the current from spin-up and 
spin-down electrons becomes $N_0\gamma_M\gamma_m$, respectively. 
In the above, it is assumed 
that 
there is no attenuation in normal metal layer so that no electron is lost 
within that layer. Since any spin independent attenuation length does not
contribute to temperature dependence of magneto current (MC) even if the 
attenuation length has temperature 
dependence. By the virtue of the fact that collector current has an exponential dependence on the electron mean free path due to the nature of hot electron transport property \cite{mean}, our assumption is acceptable when we even explore temperature 
dependence of magneto current. 
If spin-flip process is 
operating by thermal spin wave emission or absorption at finite 
temperature, the current, in parallel case, from spin-up source electrons 
can be 
calculated in the following way. $N_0\gamma_M$ electrons penetrate the 
first ferromagnetic metal layer. Among these electrons, $N_0\gamma_M(1-P(T))$ electrons
 keep their spin-up state, and $N_0\gamma_M(1-P(T))\gamma_M$ electrons will be
collected with spin-up state. However, $N_0\gamma_MP(T)$ electrons are
created with opposite spin resulting from spin-flip process, 
and $N_0\gamma_MP(T)\gamma_m$ electrons are 
collected with the spin-down. Finally,  the total number of collected 
electrons  from spin-up 
source electrons with electron number $N_0$ become 
$N_0\{\gamma_M^2(1-P(T))+\gamma_M
\gamma_mP(T)\}$. One can follow the the same scheme to calculate the
contribution to the current from spin-down source electrons. 
$N_0\gamma_m$ 
electrons penetrate the first layer, then $N_0\gamma_m(1-P(T))\gamma_m$
electrons are collected with spin-down. Meanwhile, $N_0\gamma_mP(T)$ 
electrons 
have  opposite spin state ( spin-up state). These now become the majority 
spin
 electrons to the second layer, and $N_0\gamma_m(1-P(T))\gamma_M$ 
electrons are collected. Then, the contribution to the current from 
spin-down source electrons become $N_0\{\gamma_m^2(1-P(T))+\gamma_M
\gamma_mP(T)\}$. Similarly, the current in the anti-parallel case becomes
$N_0\{\gamma_M\gamma_m(1-P(T))+\gamma_M\gamma_MP(T)\}$, and 
$N_0\{\gamma_M\gamma_m(1-P(T))+\gamma_m\gamma_mP(T)\}$ from spin-up and 
spin-down source electrons, respectively. As mentioned above, $P(T)$ 
describes spin-flip probability by thermal spin wave emission or absorption
 at finite temperature, which 
is assumed to be $P(T)=cT^{3/2}$. Here $c$ is a parameter, and $P(T)\le 1$ 
should be satisfied for any temperature T.  In our calculations we limit the temperature
ranges from zero to room temperature (T=300 K). With this limitation we 
write the spin flip probability $P(T)$ in another way. If we assume finite 
spin flip probability at room temperature (T=300 K), expressing 
$P_r$, the parameter c in $P(T)$ can be written as 
$c=P_r\times[\frac{1}{300 K}]^{3/2}$. We then write the $P(T)$ as
$P(T)=P_r\times[\frac{T}{300 K}]^{3/2}$
 
Now, the total collector current in parallel case 
influenced by spin-flip process due to thermal spin wave emission and 
absorption becomes
\eb
I_c^P=N_0\gamma_M^2[\{1+(\frac{\gamma_m}{\gamma_M})^2\}(1-P(T))
+2(\frac{\gamma_m}{\gamma_M})P(T)]. 
\ee
Similarly, in the case of anti-parallel
\eb
I_c^{AP}=N_0\gamma_M^2[\{1+(\frac{\gamma_m}{\gamma_M})^2\}P(T)
+2(\frac{\gamma_m}{\gamma_M})(1-P(T))]. 
\ee
With the expression of collector current, one can readily obtain 
the magneto current (MC) defined 
\cite{finite} such as
\eb
MC=\frac{I_c^P-I_c^{AP}}{I_c^{AP}}
\ee 
As mentioned earlier, one can easily relate phenomenological parameter
$\gamma_{M(m)}$ to hot
electron spin polarization $P_H(T)$ in such a way
\eb
\frac{\gamma_m}{\gamma_M}=\frac{1-P_H(T)}{1+P_H(T)}
\ee

Generally speaking, hot electron spin polarization will be temperature 
dependent. This implies that $\gamma_{M(m)}$ is also 
temperature 
dependent. It also will be very interesting to explore the magneto current at finite temperature due to temperature dependence of hot electron spin polarization. Relating with this issue, as remarked earlier, we do not have enough information of hot electron spin polarization at finite temperature. Therefore, We only test very simple case such as $P_H(T)=P_0(1-(T/T_c)^{3/2})$. Here, $P_0$ is spin polarization of hot electron at T=0, and $T_c$ is critical temperature of ferromagnetic metal. If one supposes that hot electron spin polarization is 
temperature independent one can obtain scaled collector 
current which is divided by $N_0\gamma_M^2$ even without knowing that prefactor. We 
express the scaled collector current as 
\eb
\tilde{I}_c^P=[\{1+(\frac{\gamma_m}{\gamma_M})^2\}(1-P(T))
+2(\frac{\gamma_m}{\gamma_M})P(T)]. 
\ee 
\eb
\tilde{I}_c^{AP}=[\{1+(\frac{\gamma_m}{\gamma_M})^2\}P(T)
+2(\frac{\gamma_m}{\gamma_M})(1-P(T))]. 
\ee  
One also
easily obtain MC. If we include temperature dependence of hot electron 
spin polarization, then we are not able to calculate collector currents 
$I_c^P$ and $I_C^{AP}$ separately because we have unknown prefactor $\gamma_M$ in Eq. (1) and Eq. (2). Fortunately, even in this
case we can still calculate temperature dependence of MC because of 
cancellation of unknown prefactor $\gamma_M$.

\section{Results and Discussion}
We now discuss the results of our model calculations. First, we explore 
the case when the hot electron spin polarization is temperature independent. In this case we assume $\gamma_{M(m)}$ is temperature independent.
Fig. 1 displays the collector current expressed in Eq. (5) with normalization at T=0.
 If there is no spin mixing effect, there is no temperature dependence as it is expected. Now, when the spin mixing process is operating one can clearly see that the collector current is decreasing with 
 increasing temperature T. 
 Fig 2 shows the collector current expressed in Eq. (6). This is relative magnitude with respect to the parallel collector current. 
 In this case, there is no temperature dependence at zero spin flip probability like parallel case. 
 However, when the spin-flip probability is 
 increasing the collector current is also increasing with temperature T.
  From these results, we find that spin-mixing effect due to  thermal spin wave at finite temperature 
contributes to the collector current quite differently depending on the relative orientation of magnetic moment in ferromagnetic metals. Fig 3. represents the temperature dependence of magneto current. As one can expect from the Fig. 1 and 2, 
we obtain that the magneto current at finite temperature accords with the experimental data in qualitative manner. 

Fig.4. displays the magneto current when we include temperature dependence of hot electron spin polarization. One can clearly see  that magneto current is decreasing even at zero spin flip probability with
increasing temperature T. This raises an interesting question. Authors of 
Ref. 8 suggests the spin-mixing effect as an origin of 
measured temperature dependence of magneto current. We also obtain 
qualitatively similar results when we consider the contribution to magneto
current  
from thermal spin wave 
emission and absorption at finite temperature without including 
temperature 
dependence of hot electron spin polarization. But, as one can see from 
Fig. 4 temperature dependence of hot electron spin polarization can 
also contribute to the behavior of magneto current at finite temperature. This fact implies that it is essential to probe relative 
importance of spin-mixing effect and temperature dependence of hot electron
spin polarization for understanding the magneto current at finite temperature in spin valve transistor. 

In conclusion, we explore the magneto current 
due to spin mixing effect from 
thermal spin wave emission and absorption at finite temperature. 
We obtain that spin mixing effect contributes to the collector current 
differently depending on the relative orientation of magnetic moment of 
ferromagnetic materials. Our calculations accords with experimental data
qualitatively. In addition, we find that temperature dependence of hot
electron spin polarization also contributes to magneto current at finite 
temperature. For clear understanding of relative importance from thermal spin wave effect and temperature dependence of hot electron spin polarization, we believe that one needs to study this from the  microscopic theory.  

\newpage
\begin{figure}
\caption{The collector current expressed in Eq. (5) with normalization at T=0. Hot electron spin polarization  $P_0$ at T=0 is taken as 
0.5, and $P_r$ represents the spin flip probability at room temperature 
(T = 300 K).}
\end{figure}
\begin{figure}
\caption{The collector current in anti-parallel case expressed in Eq. (6). This is relative magnitude with respect to the parallel current which is normalized at T=0. }
\end{figure}
\begin{figure}
\caption{Magneto current at finite temperature with temperature independent 
hot electron spin polarization.}
\end{figure}
\begin{figure}
\caption{Magneto current at finite temperature with temperature dependent
hot electron spin polarization. The form of temperature dependence is 
described in the text. Here, we take the critical 
temperature $T_c=650 K$ simulating pseudo permalloy.}
\end{figure}
permaloy
\end{document}